# Effectiveness of machining equipment user guides: A comparative study of augmented reality and traditional media

GHOBRIAL Mina[1,3,a*], SEITIER Philippe[1,b], LAGARRIGUE Pierre[3,c], GALAUP Michel[2,d], and GILLES Patrick[1,de]

[1]ICA, Institut National des Sciences Appliquées de Toulouse, 135 Av. de Rangueil, 31400 Toulouse, France

[2]EFTS, Institut National Universitaire Champollion, Pl. de Verdun, 81000 Albi, France

[3]ICA, Institut National Universitaire Champollion, Pl. de Verdun, 81000 Albi, France

[a]ghobrial@insa-toulouse.fr, [b]seitier@insa-toulouse.fr, [c]pierre.lagarrigue@univ-jfc.fr, [d]michel.galaup@univ-jfc.fr, [e]gilles@insa-toulouse.fr

**Abstract.** In the rapidly evolving landscape of manufacturing and material forming, innovative strategies are imperative for maintaining a competitive edge. Augmented Reality (AR) has emerged as a groundbreaking technology, offering new dimensions in how information is displayed and interacted with. It holds particular promise in the panel of instructional guides for complex machinery, potentially enhance traditional methods of knowledge transfer and operator training. Material forming, a key discipline within mechanical engineering, requires high-precision and skill, making it an ideal candidate for the integration of advanced instructional technologies like AR. This study aims to explore the efficiency of three distinct types of user manuals—video, paper, and augmented reality (AR)—on performance and acceptability in a material forming workshop environment. The focus will be on how AR can be specifically applied to improve task execution and understanding in material forming operations. Participants are mechanical engineering students specializing in material forming. They will engage in a series of standardized tasks related to machining processes. Performance will be gauged by metrics like task completion time and error rates, while task load will be assessed via the NASA Task Load Index (NASA-TLX) [1]. Acceptability of each manual type will be evaluated using the System Usability Scale (SUS) [2]. By comparing these various instructional formats, this research seeks to shed light on the most effective mediums for enhancing both operator performance and experience.

**Introduction**

The intersection of instructional technology and industrial training, particularly in the context of material forming, is an area of increasing interest for the industry and educational institutions. Immersive technologies—Augmented Reality (AR), Virtual Reality (VR), and Mixed Reality (MR)—particularly when applied to material forming and mechanical engineering, offer distinct advantages in enhancing traditional teaching methods. AR overlays digital information onto the real world, VR creates a fully immersive digital environment, and MR combines elements of both AR and VR, enabling interaction between virtual and real objects, based on Gandolfi's definition [3]. This study examines various types of instructional media to aid operators in efficiently using CNC machining equipment. There is a rich history of employing instructional media in industrial environments, but the emergence of immersive technologies like AR has altered the field. These advancements present unique opportunities for enhancing traditional teaching methods, especially in precision-demanding sectors like material forming in mechanical engineering. This paper seeks to explore these dynamics further, using MR (Fig. 2-a). Existing literature suggests that the adoption of immersive technologies like Augmented Reality and Mixed Reality in industrial and educational settings offers substantial benefits in terms of enhancing learning experiences and operational efficiencies. These technologies have been found to improve visualization, control, and understanding of complex manufacturing systems, contributing to more effective training and skill development. AR, in particular, enables a more efficient execution of instructions by attaching virtual objects to real-world objects, streamlining the learning process. This feature is especially beneficial in fields such as material forming and CNC machining, where the integration of instructions with the physical workspace can significantly enhance the clarity and speed of task completion, leading to a more intuitive and effective training experience. This study begins with a comprehensive review of the existing literature, shedding light on the integration and efficacy of AR and other conventional instructional media in both industrial and educational settings. The experimental approach is then detailed, which involves mechanical engineering students undertaking a task on a CNC milling machine, guided by three distinct types of instructional media. The performance of these tools is assessed through various metrics.

**Literature Review**
J. Butt [4] discusses the evolution and integration of additive manufacturing and Augmented Reality technologies. Stereolithography in the 1980s, focused on rapid prototyping using various materials. Over time, it expanded towards mass-scale production. AR technology, evolving since the 1960s, superimposes computer-generated images in real environments, enhancing user interaction. AR and additive manufacturing have increasingly intersected for AR's capability to superimpose virtual instructions onto physical workspaces, which significantly enhances task clarity and execution speed especially for complex material forming machines. Gonzalez-Franco *et al.* [5] explore the effectiveness of AR in complex manufacturing training, specifically in aircraft maintenance. Their study, involving 20 participants, compares immersive AR training with conventional face-to-face training methods. The research, utilizing a Mixed Reality setup with Head-Mounted Displays, found no significant difference in performance levels between the two training methods. This study underscores the potential of AR as a viable alternative to traditional training, offering insights into the application of collaborative Mixed Reality in transmitting procedural knowledge. In the context of additive manufacturing, the research by Ostrander *et al.* [6] explores the use of VR as an instructional tool. They investigate the effectiveness of VR in teaching introductory additive manufacturing concepts, comparing interactive and passive VR lessons. Their findings indicate that both forms of VR instruction can effectively convey technical concepts, with interactive VR showing advantages in enhancing self-efficacy. This study highlights the versatility of VR in educational settings and its potential in overcoming the limitations of traditional industrial training methods. Additionally, Mogessie *et al.* [7] address the demand for training in metals additive manufacturing machines, focusing on the EOS M290. Their approach involves an interactive VR training system, the 'AM Training Tutor', which demonstrates the utility of VR in specialized training scenarios. The ongoing development of a modular and generic version of this system suggests a future where VR training can be easily adapted for various machines, highlighting the scalability and customization potential of VR-based instructional technologies. A study by Botto *et al.* [8] focuses on the impact of AR on manual assembly operations in the manufacturing industry. Their research shows that AR-based assembly assistants can support key activities such as identification, handling, alignment, joining, adjustment, and inspection. They developed a tablet-based AR tool that integrates these activities and compared its effectiveness against traditional paper-based instructions. The study reveals that the AR tool generally reduces errors but increases the time needed for assembly, suggesting a trade-off between accuracy and efficiency [8]. The article by Sitko *et al.* [9] adds a new dimension to the existing literature by focusing specifically on the forging industry. It provides a comparative perspective between VR and AR, highlighting AR's practical applications and benefits in an industrial setting that faces unique challenges. This broadens the scope of this review by linking

AR not only to training and manufacturing but also to maintenance and operational efficiency in heavy industries. The work of Seitier *et al.* [10] focused onto the implementation of AR in mechanical engineering education, exploring its potential to enhance learning experiences. Meanwhile, Ghobrial *et al.* [11] conducted a specific study on the use of AR with a 3D printer as a simplified Additive Manufacturing machine, assessing its acceptability and success rate in an industrial context. This study builds upon previous works by adding a new dimension to the use of AR in mechanical engineering for both industrial and educational purposes. It conducts a comparative analysis throughout various study axes, offering an in-depth evaluation of how AR improves training performance. This contributes considerably to the existing knowledge on instructional technology in industrial and educational training.

**Experimental Setup**
In this comparative study, the chosen procedure involves loading a new tool into a HURON VX8 machining center (Fig. 1). This process consists of three main stages, each involving a series of elementary actions (text entry, button pressing, door opening, etc.). The three main stages are the creation of the tool file, its allocation to a slot in the tool magazine, and the loading of the tool into the machine's tool slot. This is a procedure where students often make mistakes, leading to machine blockages. Therefore, developing an effective instructional support, meaning one that minimizes handling errors, is important. The experimental setup was carefully designed to evaluate the effectiveness of different educational supports (paper, video and Augmented Reality) to guide students during the loading of a new tool. Mechanical engineering students participated in this study. The setup involved the following key elements: *(i)* Participants were arbitrarily assigned to one of three stations, each one receiving a different type of instructional media: paper manual, video guide, or a AR headset equipped with an interactive scenario, *(ii)* before beginning the task, students completed a preliminary form providing general information such as their name, age, year in their mechanical engineering program, previous experience with CNC machines, and any prior experience with Head-Mounted Displays such as Virtual Reality headsets, *(iii)* the task assigned to each student was a tool load operation on a VX8 3-axis CNC machine. This procedure was chosen for its relevance to the usage of a typical CNC machine that is common in mechanical engineering discipline and has a specific protocol to each machine, *(iv)* a specific indication was given to the students, instructing them to seek the supervisor's help only if they were completely blocked at any step of the procedure. The supervisor was also responsible for intervening if a student was engaged in any action that could be dangerous or potentially damaging to the equipment. The frequency of such interventions was recorded for each participant, *(v)* the timing for each student's task was recorded. The clock started as soon as the participant finished filling out the initial form and stopped upon the completion of the tool load procedure.

 A total of 96 mechanical engineering students, aged between 18 and 24 years old, participated in the experiment. These students represented various stages of their academic journey, ranging from the first year to the fourth year of their program. Students were evenly distributed across the different instructional media. This diverse and balanced participation ensured a comprehensive assessment of each instructional medium's effectiveness. Additionally, to ensure the integrity of the study's results, each student participated in the experiment only once, using one type of instructional media. This approach prevented any learning or familiarity effects from repeated exposure to the task, thus maintaining the validity of the comparative analysis between the different instructional media.

**Instructional media**
The three instructional media used in this study were carefully selected and designed to offer a diverse range of learning experiences: *(i)* The paper guide consisted of the official constructor's user guide for executing the tool load task. The guide provided step-by-step instructions, along with diagrams and safety precautions. It was intended to replicate a traditional learning method, where students rely on written documentation to understand and perform the procedure, *(ii)* the video guide consists of a comprehensive video was prepared by the shop's teachers and technicians, demonstrating the same tool load procedure as outlined in the paper manual. This video, displayed on a tablet at the respective station (Fig. 1), allowed students to visually follow the process in a sequential manner. The video aimed to provide a more dynamic and engaging learning experience compared to the static paper guide, potentially catering to students who learn better through visual demonstrations.

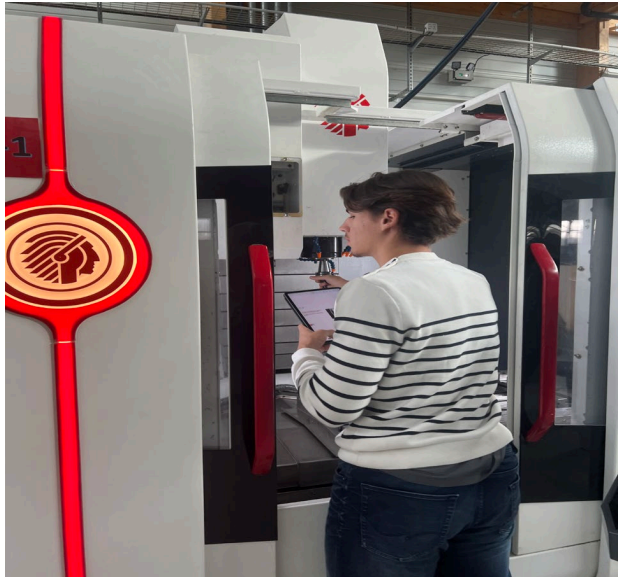

*Figure 1 - Student using video support*

*(ii)* For the Augmented Reality guide, a Microsoft Hololens 2 headset was used (Fig. 2-a), the integrated application was developed using Unity, a development platform. The AR guide creation process does not solely depend on Unity, some alternatives such as Unreal Engine could serve as a development solution. Vuforia Studio is another alternative that provides an off-the-shelf solution for AR development. Diota is an additional option that could be explored, offering AR tools for industrial sectors. The AR guide featured virtual objects that were superimposed onto the real-world environment of the CNC machine (Fig. 2-b) and a hand-attached menu (Fig. 2-c), in each step of the procedure. Some 3D models were integrated into the virtual scene. While pointing arrows were sourced from the Unity Asset Store, which offers a wide range of ready-made models and prefabs, other models were obtained directly from the makers, such as the Huron VX8 CNC machine's digital model. The tool holder, was custom-modeled using 3DExperience software to match the used tool holders. The hand-attached menu and hand coaches, which are interactive tools designed to facilitate user interaction within the AR environment, utilized Mixed Reality Toolkit (MRTK) prefabs. MRTK is a collection of scripts and components intended to accelerate the development of mixed reality applications. This interactive scenario was designed to guide students through the tool load process in an immersive way. Additionally, voice instructions were integrated to assist students who might have difficulty reading or following written instructions, due to the current state of the art of the AR hardware.

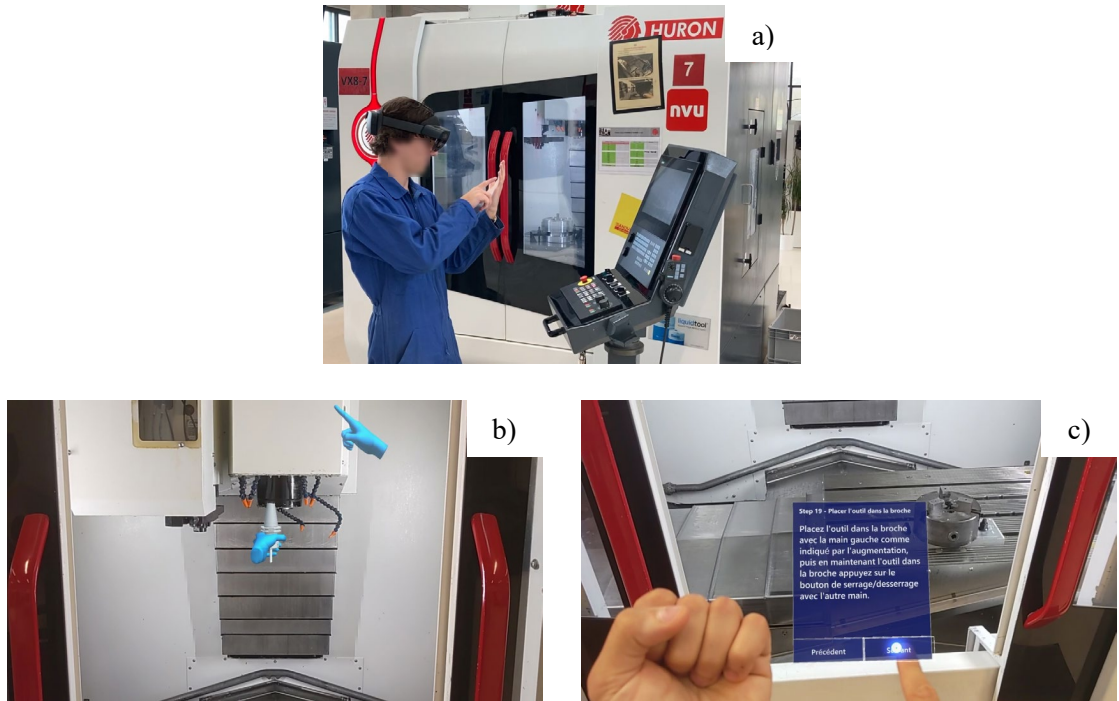

*Figure 2 - Student using AR support*

Each of these instructional media was designed to cater to different learning tools and to evaluate the effectiveness of traditional media versus Augmented Reality in a mechanical engineering setting. The comparative study aims to assess not only the students' acceptability of each medium but also the students' efficiency, autonomy and comfort levels with each type of learning resource.

**Results and discussion**
The performance outcomes of the three instructional tools is analyzed, emphasizing their impact on user engagement and productivity. A range of established metrics is utilized to assess aspects such as usability, task load, dependence, and task execution time. These metrics serve as a quantitative basis for comparison and shed light on the subjective user experiences. The examination of these results provides insights into the advantages and drawbacks of each tool. The following part details the methodologies behind each metric and the interpretation of the data: (i) System Usability Scale (SUS): The SUS is a simple, ten-item scale giving a global view of subjective assessments of usability. Each item is scored on a scale of 1 to 5, where 1 represents the lowest usability and 5 represents the highest, with alternating positive and negative connotations for the questions in order to keep the users' attention throughout the questions. The scores are converted to a 0-100 scale for easier interpretation, with 100 representing the best possible usability. Thus, a higher SUS score indicates better usability, (ii) NASA Task Load Index (NASA-TLX): The NASA-TLX is a widely-used, subjective workload assessment tool that rates perceived workload in order to assess a task, system, or team's effectiveness or other aspects of performance. It measures workload on a scale from 1 to 20, where 1 indicates no workload and 20 indicates an extremely high workload. Contrary to SUS, a higher score on the NASA-TLX indicates a higher load and, typically, a more negative impact on performance, (iii) dependence in this context is measured by the number of times a supervisor intervened. Fewer interventions indicate that the user can work more independently, which is considered better, (iv) task execution time is measured in order to evaluate efficiency, the lower the time required to execute the task, the more efficient the process. The paper guide demonstrates the highest dispersion in usability (Fig. 3-a) and task execution time (Fig. 3-b), suggesting inconsistent performance among users. Its lower peak in all metrics may indicate technological limitations, such as the lack of interactivity and real-time support, which could impede user performance and learning. The video guide shows a narrower dispersion, particularly in usability (Fig. 3-a) and dependence (Fig. 3-d), indicating more consistent performance. Its higher peaks suggest that video is more user-friendly and supports greater independence, possibly due to clear visual instructions and the ability to replay content. AR presents a significant time efficiency advantage with a peak similar to video, but with wider dispersion. This could imply that while AR can be highly effective, its performance fluctuates considerably among users, potentially due to ergonomic challenges or inconsistent technological maturity of AR devices. Its higher peak in dependence, relative to video, suggests that the used AR device may require more user effort or support to complete tasks independently.

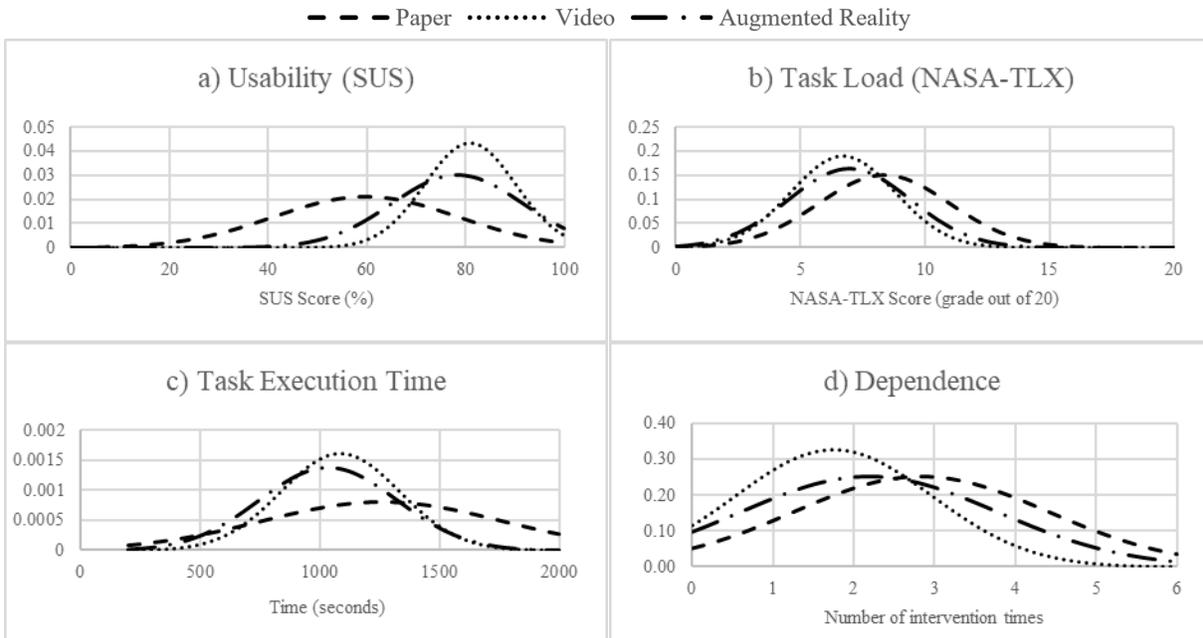

*Figure 3 – Usability (a), Task Load (b), Task Execution Time (c) and Dependence (d) normal distributions*

In order to establish a basis for comparison, all metrics but SUS scores were recalculated as percentages, with higher values indicating more positive tool performance. Here's an explanation of how each of these indicators are calculated: *(i)* System Usability Scale (SUS): The scores are converted to a 0-100 scale for easier interpretation, with 100 representing the best possible usability. Thus, a higher SUS score indicates better usability. Similarly, other indicators have been recalibrated to the same 0-100 scale to ensure uniformity and ease of comparison, *(ii)* NASA Task Load Index (NASA-TLX): For this indicator, the scores have been inverted and converted to a 0-100 scale for consistency in presentation, the metric will be called Task Ease (Eq. 1).

$$Task\ Ease\ \% = (1 - \frac{NASA-TLX\ score - 1}{19}) \times 100 \qquad (1)$$

*(iii)* This indicator was calculated by normalizing the actual number of times the supervisor intervened (dependence) against the range of observed number of times, with the lowest recorded number of times set as a reference point for 100% autonomy and the highest recorded number of times set as a reference point for 0% autonomy (Eq. 2).

$$Autonomy\ \% = \left(1 - \frac{Number\ of\ times\ of\ intervention}{Highest\ Number - Lowest\ Number}\right) \times 100 \qquad (2)$$

*(iv)* To create a consistent indicator, task execution time will also be converted to a positive percentage metric, with 100% being the most efficient. This percentage was also calculated by normalizing the actual task times against the range of observed times, with the lowest recorded task time set as a reference point for 100% efficiency and the highest recorded time set as a reference point for 0% efficiency (Eq. 3).

$$Efficiency\ \% = \left(1 - \frac{Task\ Time - Best\ Time}{Worst\ Time - Best\ Time}\right) \times 100 \qquad (3)$$

By inverting Task Load scale and normalizing the others, the study presents all the metrics on a consistent scale where a higher percentage always indicates a better outcome. This approach simplifies the comparison of different instructional tools across various performance metrics. The radar chart (Fig. 4) visually encapsulates the reported findings, plotting the performance of paper, video, and AR tools along axes that represent key metrics: Usability (as measured by the System Usability Scale), Task Ease (derived from the NASA-TLX score), Efficiency, and Autonomy. On the chart, video appears to offer a balanced performance across all metrics, scoring particularly high in usability and autonomy, indicating it is user-friendly and supports independent learning. Augmented Reality shows a notable efficiency advantage, potentially due to its interactive and engaging nature. However, it does not reach the same level of performance in autonomy, which may reflect the current ergonomic and technological maturity issues previously discussed. Paper scores the lowest across most metrics, particularly in efficiency and usability, suggesting that it is the least effective tool among the three for this application.

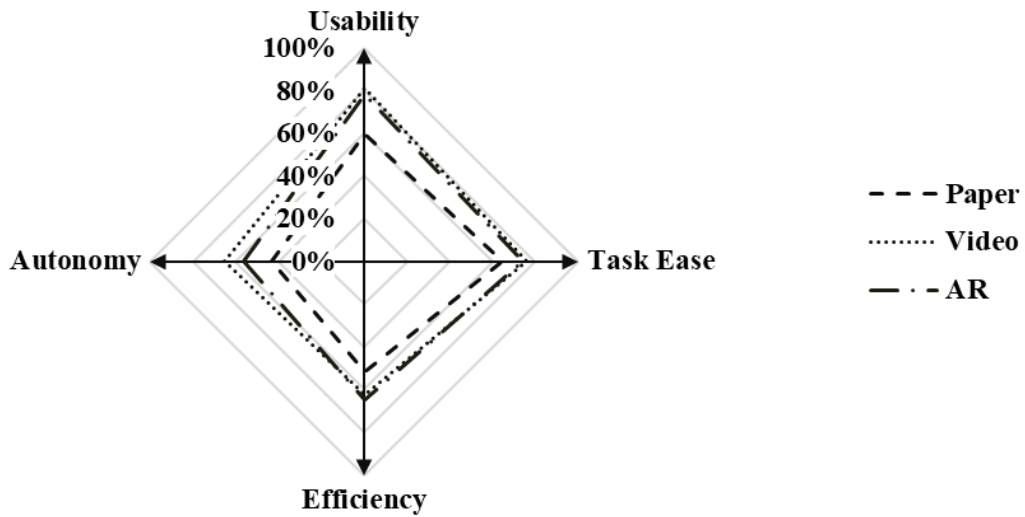

*Figure 4 - Tool Performance Radar Representation*

The findings suggest that Augmented Reality has the potential to enhance efficiency in learning environments. However, students that were assigned the AR headset reported several issues that may affect its practicality. These included difficulty with visibility due to lighting conditions and challenges with focus and accommodation, making reading instructions problematic for some. Additionally, the physical weight of the headset was noticeable to some users. The most common feedback related to hand tracking and interaction with the system's buttons; students familiar with VR or who adapted quickly experienced fewer issues, highlighting a learning curve and adaptation factor in the effective use of AR for educational purposes. These interpretations align with the discussed results, indicating that while AR is promising for efficiency, video guides currently offer a more user-friendly and autonomous experience, and paper guides lag behind in these metrics, likely due to their static nature and lack of interactive features.

**Conclusion**

The results further emphasize that the Augmented Reality technology, while promising, may still be facing challenges due to its current level of technological maturity. The efficiency advantage of AR as indicated in the graphs suggests it has significant potential for instructional use. However, the higher number of supervisor interventions required as compared to video guides could be indicative of ergonomics-related problems with the current state-of-the-art AR hardware in this application. These ergonomic challenges could stem from issues like user discomfort due to the weight or fit of the AR headset, interface complexity, or limitations in the user interaction design, which can hinder the learning process and require additional external guidance. Moreover, the lack of technological maturity might manifest in the form of less intuitive user experiences or technical reliability issues, which could explain why AR did not score as well as video in terms of autonomy. To summarize, the results suggest that while AR is an efficient medium, it may not yet provide the seamless and independent learning experience that more mature technologies like video can offer due to ergonomic and maturity-related limitations. These findings would likely be in line with discussions in the article's literature review, which have touched upon the current technological and ergonomic barriers that AR needs to overcome to achieve broader adoption and effectiveness in educational settings. It's important to note that AR technology, while showing promise, must undergo further exploration and testing using diverse mediums and devices. Further applications of AR in mechanical engineering training will be tested. This includes, on one hand, the creation of new applications for other types of machinery such as tool measurement benches, and on the other hand, the transposition of existing AR applications into tablet-based AR. In particular, the integration of AR in tablet mediums may effectively blend the advantages of video and AR, potentially offering an optimized balance of user-friendliness and interactive efficiency. This approach aims to harness the strengths of each medium to enhance the overall learning experience.